\documentclass[aps,pra,superscriptaddress,preprint,onecolumn,showpacs,amsmath,amssymb,floatfix,nofootinbib]{revtex4}
\usepackage{amsthm,graphicx,wasysym,mathtools}

\newcommand{\la}{\lambda}
\newcommand{\lo}{\lambda_0}
\newcommand{\lao}{\overline{\lambda}}
\newcommand{\s}{\sigma}
\newcommand{\al}{\alpha}
\newcommand{\e}{\epsilon}
\newcommand{\Lper}{\mathcal L}

\newcommand{\rb}{{\bf r}}

\newcommand{\phir}{|\Phi\rangle}

\newcommand{\n}[1] { |\vec n_{#1}\rangle }
\newcommand{\G}{\Gamma}

\begin{document}
\title{The attractive Bose gas in two dimensions: an analytical study of its fragmentation and collapse}
\author{Marios C. Tsatsos\footnote[2]{Temporary address: Instituto de F\' isica, Universidad Nacional Aut\' onoma de M\' exico. Apartado Postal 20-364, 01000 M\' exico D. F. Mexico. Past address: Theoretische Chemie, Physikalisch-Chemisches Institut, Universit\"at Heidelberg, Im Neuenheimer Feld 229, D-69120 Heidelberg, Germany. \\ Electronic address: \tt{mariostsatsos@gmail.com}}
}
\affiliation{Instituto de F\' isica de S\~ao Carlos, Universidade de S\~ao Paulo, Caixa Postal 369, 13560-970 S\~ao Carlos, S\~ao Paulo, Brazil}
\date{\today}

\begin{abstract}
An attractive Bose-Einstein condensate in two spatial dimensions is expected to collapse for supercritical values of the interaction strength. Moreover, it is known that for nonzero quanta of angular momentum and infinitesimal attraction the gas prefers to fragment and distribute its angular momentum over different orbitals.
In this work we examine the two-dimensional trapped Bose gas for finite values of attraction and describe the ground state in connection to its angular momentum by theoretical methods that go beyond the standard Gross-Pitaevskii theory. By applying the \emph{best mean field} approach over a variational ansatz whose accuracy has been checked numerically, we derive analytical relations for the energy, the fragmentation of the ground states and the critical (for collapse) value of the attraction strength as a function of the total angular momentum $L$. 
\end{abstract}

\maketitle
\thispagestyle{empty}

\section{Introduction}
One unique feature of ultracold trapped atomic gases is the deterministic control that can be exerted -- in the laboratory -- on the gas and its properties. The density of the gas, the geometry of the trap, as well as the strength and sign of interaction have nowadays become fully controllable. Deeper theoretical understanding and advanced experimental tools allow one to change the sign of the interaction between the bosonic atoms of the gas in a controllable manner and attractive Bose-Einstein condensates (BEC) can be formed in the laboratory \cite{Inouye1998,Roberts1998,Courteille1998}.
The novel collapse phenomenon, which is absent in the repulsive gas, has given the attractive BEC a special position in contemporary research and manifested itself in the fascinating colllapse and `Bosenova' experiments  \cite{Experiment_Collapse, Controlled_Collapse}.

Not accidentally, one of the first case-studies of occurrence of fragmentation in BECs was that of an attractive boson gas in two spatial dimensions \cite{Wilkin1998}. There the authors described the bosonic gas with a many-body ansatz that appears to be successful at least in the limit where the interparticle interaction is very weak. One important result of this work was to find the natural orbitals and their natural occupations in a simple analytic expression. Based on this, it was derived that for a given angular momentum $L$ the ground state of the system is fragmented. In other words, a non-vanishing angular momentum of the system causes the bosons of the gas to be distributed over a vast number of single-particle states, rather than one. Thus, coherence is lost and this renders the system \emph{not} describable by the standard Gross-Pitaevskii theory. In response to this finding it has been suggested that the definition of the single-particle reduced density matrix and the definition of Bose-Einstein condensation should be modified \cite{Pethick2000} or that in the absence of the symmetry (isotropy) of the trapping potential the fragmentation will vanish \cite{Jackson2008}. However, these do not clash with a main characteristic of the attractive gas: the angular momentum $L$ is imprinted in the gas in a completely different -- fragmented -- way, than that in the repulsive case.

Still, the non-weakly attractive two-dimensional gas and its collapse has not been scrutinized in the light of the above findings. In three spatial dimensions on the other hand, it has been shown that fragmentation and participation of the low-lying excited states \cite{Cederbaum2008} and the presence of total angular momentum $L$  \cite{Tsatsos2010} can postpone the collapse. 
In the present work we examine the structure of the ground state of finite systems with non-zero angular momentum (AM) and finite non-weak interaction strength $\la$. We express the energy of this ground state (GS) as a function of $L$ and, moreover, we find an expression for the critical (maximum allowed) value $\la_c$ of the interaction strength. The method used is the best mean field (BMF) theory, that has been introduced and described in Ref.~\cite{Streltsov2003}. The orbital basis consists of modified Gaussian orbitals (scaled single-particle states of the harmonic oscillator). Using that, we reveal the structure of the ground state with $L>0$: it is a distribution of the bosons over the $M$ orbitals that above some $L$, differs from the one derived in the many-body (MB) treatment \cite{Wilkin1998}. However, the energy that we derive for this state can drop lower than that of the GS of the abovementioned work (also others, see for example Ref. \cite{Jackson2008}). Asymptotically, in the limit of very weak interaction strength $\lambda$ and large particle number $N$, our expression gives back the previously known one.

The structure of this paper is the following. We introduce the Hamiltonian of the system and the mean field (MF) ansatz in Sec. \ref{2D_system_intro}. In Sec.~\ref{2D_GS_energy} we derive an expression for the energy of this ground state as a function of the AM $L$, for any finite $L$ and non-weak $\la$. We show that our expression encompasses the energy known from previous asymptotic MB and MF results. Additionally and in connection to this finding, in Sec.~\ref{2D_critical_la} we derive an expression for the critical value of the interaction strength as a function of the AM $L$. In Sec.~\ref{MBcomparison} we compare our results to previously known ones. Lastly, we conclude and discuss the findings in Sec.~\ref{conclusions}.

\section{The system \label{2D_system_intro}}
We consider the Hamiltonian $ H=H_0+\lo W$ with
\begin{equation}
H_0=\frac{1}{2}\sum_i^N \left(-\nabla_{\rb_i}^2+\rb_i^2\right)~\text{and}~W=\sum_{i<j}^N \delta(\rb_i-\rb_j),
\end{equation}
in dimensionless units where $\omega=\hbar=m=1$. For $\lo<0$ the above Hamiltonian describes a 2D trapped gas of attractive ultracold bosons.
To represent the wave function of the system we use the general MF ansatz (Fock state): 
\begin{equation}
|\Phi\rangle= {\mathcal S} \phi_1(\rb_1)\dots\phi_1(\rb_{n_1}) \phi_2(\rb_{n_1+1})\dots\phi_2(\rb_{n_2})\dots\phi_M(\rb_N) \equiv |n_1,n_2,\dots,n_M\rangle,
 \label{permanent}
\end{equation}
where ${\mathcal S}$ is the symmetrizing operator, accounting for the bosonic nature of the wave function. The ansatz of Eq.~(\ref{permanent}) describes a fragmented system of $N$ particles, where $n_i$ of them reside in the $\phi_i$ single-particle state (orbital), with $i=1,\dots,M$, $\sum_i^M n_i=N$. The total density (i.e., diagonal of the single-particle reduced density matrix) of this state is $\rho(r)=\sum_{i=1}^M n_i |\phi_i(r)|^2 $. The total (expectation value of) angular momentum of this state is $L=\sum_i l_i n_i$, where $l_i=\langle \phi_i|\hat L_z|\phi_i\rangle$ is the orbital angular momentum of the orbital $\phi_i$. Generally, $l_i$ is a function of the anisotropy of the trap. In the case of an isotropically trapped gas, that we examine here, the orbitals are expected to be eigenstates of the (single-particle) operator $\hat L_z(\rb)$ and hence the expectation values $l_i$ equal the eigenvalues $l_i=0,1,2,\dots$~. 
Evaluated on the above ansatz [Eq.~\eqref{permanent}] the total energy takes on the appearance \cite{Streltsov2003,MCHB}:
\begin{equation}
E= \langle\Phi |H|\Phi\rangle = \sum_i^M \left(\rho_i h_i + \frac \lo 2 \rho_{ii} w_{i,i} + \lo\sum_{j\neq i}^M \rho_{ij} w_{i,j}\right),
\label{En_Phi}
\end{equation}
where $h_i=\langle \phi_i|H_0|\phi_i\rangle$, $w_{i,j}=\langle \phi_i\phi_j|\phi_i\phi_j\rangle$ and $\rho_i=n_i, \rho_{ii}=n_i^2-n_i$, and $\rho_{ij}=n_i n_j$ are the diagonal matrix elements of the single- and two-particle densities.
The task of the present work is to find the best mean field: the configuration of Eq.~\eqref{permanent} that corresponds to the lowest possible energy.

To represent the single-particle states or orbitals $\phi_i$, that the bosons of the system occupy, we use the Gaussian solutions of the 2D harmonic oscillator \cite*{Fetter} that have been parametrized with a variational parameter $\s$. The variation of this parameter can capture the contraction and collapse of the gas due to the attraction. This variational approach has been, in the past, scrutinized and compared to numerical solutions and found to provide a satisfactory approximation to the ground \cite{Duine2001} as well as non-ground states \cite{Cederbaum2008} of the attractive gas. Moreover, current preliminary numerical analysis suggests that the GS of the gas indeed has a Gaussian-like profile for all allowed interaction strengths, so that $\sigma$-scaled Gaussian modes is a justified approximation.
In the following analysis we choose two different but related orthonormal orbital subsets. At first, we make use of the orbital basis $\{\phi_{lm}\},~m=-l,-(l-2),\dots, l-2,l$ consisting of the $s,p,d$ and $f$-type orbitals that solve exactly the 2D non-interacting problem. The orbitals are scaled by a parameter $\s$ which is to be found variationally and hence optimizes the width of the Gaussian. This particular scaling does not affect the orbital angular momentum (OAM) $\{m\}=\{0,1,-1,2,0,-2, \dots \}$ that the orbitals carry, i.e., they are still eigenfunctions of $\hat L_z(\rb)$. Then, and in order to include higher AM, we switch to the basis consisting of the single-particle functions with quantum number $m=l$ only. This basis, which is also referred to as the \emph{lowest Landau levels (LLL)} \cite{Wilkin1998}, is explicitly written as:
\begin{equation}
 \phi_m(\rb)=N_m \left(\frac{r}{\s}\right)^m e^{-r^2/(2\s^2)} e^{i m\theta},
\label{LLL}
\end{equation}
where $N_m=(\pi \s^2 m!)^{-1/2}$ is the normalization constant and $\s > 0$ the scaling parameter. 
Thus, picking up states only with $m=l$ makes the latter set (LLL) a subset of the former one $\{s,p,d,f \dots\}$. We demonstrate in the following that the BMF for a given non-zero total AM $L$ is the state that includes the LLL only. That is, a variational calculation of the energy of a state built over the orbitals of a general $\{s,p,d,f\dots\}$-basis yields zero occupation numbers for the single-particle states that do not belong to the LLL (non-LLLs). Furthermore, we show that the $L=0$ ground state, for any number of orbitals, is a condensed coherent state, while a generic $L>0$ state is in principle energetically favorable if it is fragmented. However, the fragmentation ratio is found not to be high.

The GS state of the attractive system is expected to collapse if the parameter $\la=|\lo| (N-1)$ exceeds a critical value $\la_c$ \cite*{Fetter, Ruprecht_collapse, Collapse-supernova}. The same holds true for excited states, with the critical value for collapse $\la_c$ now shifted to higher values \cite{Adhikari2001, Tsatsos2010}. Here, the inclusion of the variational scaling parameter into the orbitals allows for a good description of the collapse of the condensate and does not constrain the discussion to the limit where $\la\ll1$, as done in \cite{Wilkin1998, Mottelson1999}, which is far from the collapse.

\section{Energy of the ground states \label{2D_GS_energy}}
By substituting the constraints $N=\sum n_i,L=\sum l_i n_i$ and using the symbols $\al_i=n_i/N$ for the relative occupation, the energy functional of Eq.~(\ref{En_Phi}) takes on the form:
\begin{align}
\e  & = E/N   =  (1+\Lper) h_{00} + \frac{\lao}{2}w_{00}\left(\frac{\Lper^2}{2}-1-\frac{\Lper-2}{2N}\right)  +   \nonumber \\ 
&+ \sum_{lm}\left\lbrace(l-m)h_{00} + \frac \lao 2 \left[\left(2-\Lper~m + \frac{m-2}{2N} \right)w_{00}+ 4 (\Lper-1)w_{00,lm}- 4 \Lper w_{11,lm}\right]\right\rbrace \al_{lm}+ \nonumber \\
&+ \frac \lao 2 \sum_{lm,l'm'} \left(\mathcal K_{lm,l'm'}^+ - \mathcal K_{lm,l'm'}^- \right) \al_{lm} \al_{l'm'}, 
 \label{Energies}
\end{align}
with $\lao=|\lo|N$, $\Lper=L/N$, $h_{00}=(1+\s^4)/(2\s^2)$, $w_{00} \equiv w_{00,00}=1/(2\pi\s^2)$,  
$\mathcal K^+ = m (\frac m 2 +1) w_{00} + 4 w_{00,l'm'}+4mw_{11,l'm'}+2 w_{lm,l'm'} (1-\delta_{lm,l'm'})$ 
and $\mathcal K^-= m w_{00}+4 m w_{00,l'm'}$ the positive and negative prefactor of the square terms 
$\al_{lm}\al_{l'm'}$ accordingly. The summations run over $-l\leq m \leq l,~0\leq l \leq M$ excluding the pairs $l=m=0$ and $l=m=1$. It should be noted that we have changed the representation from $n_i$ to $\al_{lm}$. The prefactors $\mathcal K^+$ and $\mathcal K^-$ depend solely on the indices $lm,l'm'$ and not on the AM $L$. It is crucial here to explicitly consider the constants of motion  $L$ and $N$ in the above expression. To see that consider a vanishing interaction, $\la=0$, or an infinitesimal one, $\la\ll1$. Then the above expression for the energy yields immediately that the optimal distribution is the one with $m=l$, that is the LLL. 
We ask: what is the optimal distribution of $\al_{lm}$ that minimizes the polynomial of Eq.~(\ref{Energies}) for some given finite $\la, L$ and $N$. To answer this, we first consider only \emph{small oscillations} of the (non-negative) occupations $\al_{lm}$ around 0. Since $0\leq\al_{lm}\leq1$, for all $l$ and $m$ we can truncate quadratic terms $\mathcal O(\al_{lm}^2)$ and $\mathcal O(\al_{lm} \al_{l'm'})$ and study the behavior of the linearized (in terms of $\al_{lm}$) energy.

\subsection{Zero Angular Momentum}
First, we focus on the states that possess no angular momentum, i.e., $L=0$. In the case of zero AM the prefactor of $\al_{lm}$ of Eq.~(\ref{Energies}) becomes  $$\sum_{lm}\left[(l-m)h_{00}+\lao\left(1+\frac{m-2}{4N}\right)w_{00}- \lao 2 w_{00,lm}\right].$$ Its first term is always non-negative ($l\geq m$) while for the integrals $w_{00,lm}$ we (numerically) found that $0\leq w_{00,lm} \leq \frac 1 2 w_{00}$, as long as $lm\neq 00$. Recalling that $\lao>0$, we see that the prefactor that multiplies $\la$ will always be positive. Hence, any non-zero value for the occupations $\al_{lm}$ (excluding $\al_{00}, \al_{11}$) will only increase the energy and thus fragmentation is not energetically favorable. That is, for all allowed $\la$ the overall GS of the system with vanishing AM is the condensed state $|\vec n_0\rangle= |N,0,\dots,0\rangle$. The energy of Eq.~(\ref{Energies}) for this GS is $\e_0=h_{00}(\s)-\frac\la 2 w_{00}(\s)$. By optimizing the latter with respect to $\s$ we end up with the expression 
\begin{equation}
\e_0=E_0/N=\sqrt{1-\frac{\la}{2\pi}},
\label{GSgpenergy}
\end{equation}
which is of course the GP energy.

\subsection{Finite Angular Momentum and Lowest Landau Levels}
We now turn to the case of non-vanishing $L$. As we shall see in this section, the presence of AM can change the picture. First we show that the minimization of Eq.~(\ref{Energies}) yields an optimal distribution of $\al$'s (or $n_i$'s) over the LLL only. We stress here that the LLL has been widely used as a basis for the description of the ground state with $L>0$ and known to be an adequate approximation \cite{Morris2006}. We provide, in addition and, to the best of our knowledge for the first time, a variational argument for the validity of the LLL. It is clear from Eq.~(\ref{Energies}) that the part of the energy not depending on $\lambda$ admits a minimum when only the $m=l$ single-particle states contribute to the energy functional. The second term linear in $\alpha_{lm}$ (with prefactor $-\Lper m w_{00}$) drops linearly with $m$ and hence minimizes the energy when $m=\text{max}=l$. For the matrix elements $w_{00,lm}$ we have noticed (up to $l=3$) that their value is minimal at $m=l$, while the opposite holds true for the $w_{11,lm}$ elements. That is, they are a non-decreasing function of $m$ (for given $l$). Taking into account the signs of each of the terms we see that the total energy functional, in a first order approximation to $\al$, admits a minimum when $m=l$. This means that only the LLL orbitals can have non-zero occupations, for non-zero total AM $L$.
We verify this behavior, i.e., that in the GS with given L only orbitals-members of the LLL are occupied, by including terms of second order as well. To do so, we first examine the energy of the state $\phir$ built over three orbitals with different AM quantum numbers. Consider the permanents
\begin{equation}
|n_0,n_+,n_-\rangle \equiv |N(1-\Lper-2\al_-), N(\Lper+\al_-), N \al_-\rangle,
 \label{permanent3}
\end{equation}
where $n_0,n_+,n_-$ are, respectively, the occupations of the $\phi_{00},\phi_{11},\phi_{1-1}$ single-particle states (or, equivalently, the $s,p_+,p_-$ orbitals) with $n_0+n_++n_-=N$, $L=n_+-n_-$ is the total AM of the state, $\Lper=L/N$ the non-negative AM per particle and $\al_-=\frac{n_-}{N}$. In this configurations the states $\phi_{00}$ and $\phi_{11}$ comprise the LLL while the $\phi_{1-1}$ orbital is a non-LLL state.
We express the total energy as a function of the occupations $n_+,n_-$ (or equivalently the parameters $\Lper,\al_-$) and the scaling parameter $\s$. By minimizing this expression with respect to $\s$ we obtain, in the large-$N$ limit, the expression for the total energy:
\begin{equation}
\e=E/N=\frac{\sqrt{1+\Lper+2\al_-} \sqrt{4\pi (1+\Lper+2\al_-) + (\Lper^2+2\Lper \al_-+2\al_-^2-2)\la}}{2\sqrt{\pi}}
\label{EnergyL}
\end{equation}
or, in the limit of weak interaction ($\la \ll 1$),
\begin{equation}
\e=1+\Lper+2\al_-+\frac{\Lper^2+2\Lper \al_-+2\al_-^2-2}{8 \pi} \la + \mathcal O(\la^2).
\label{EnergyL_app}
\end{equation}
It is easily seen in the last two equations that any non-zero value of the parameter $\al_-$ will only increase the total energy and this demonstrates that the non-LLL orbital (here $\al_-$) is not energetically favored for a given $L>0$. The above expressions for the energy are given for brevity in the presentation in the large-$N$ limit only. However, the situation is not different if one considers the full expression.

\begin{figure}[ht]
 \includegraphics[scale=0.6]{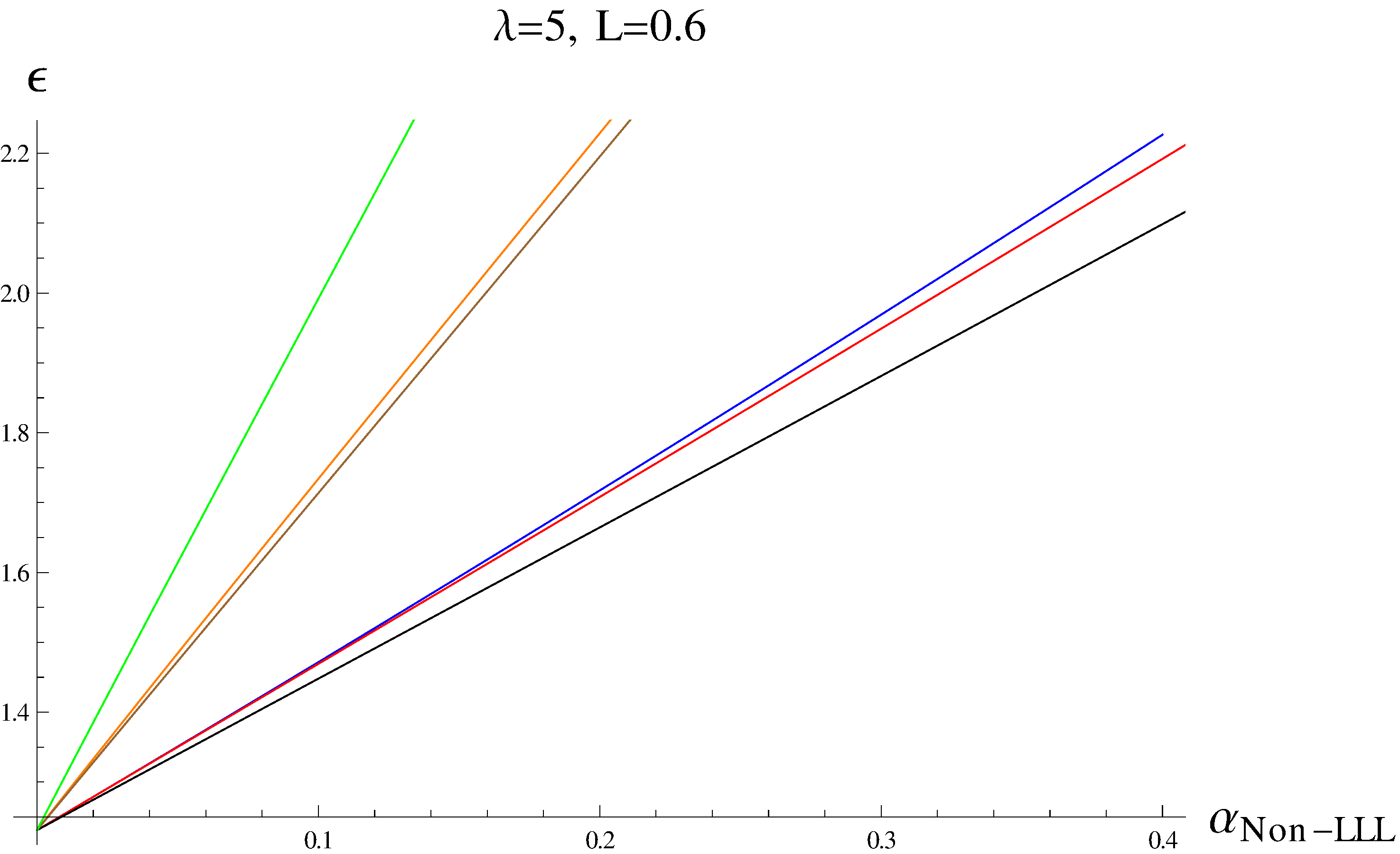}
 \includegraphics[scale=0.6]{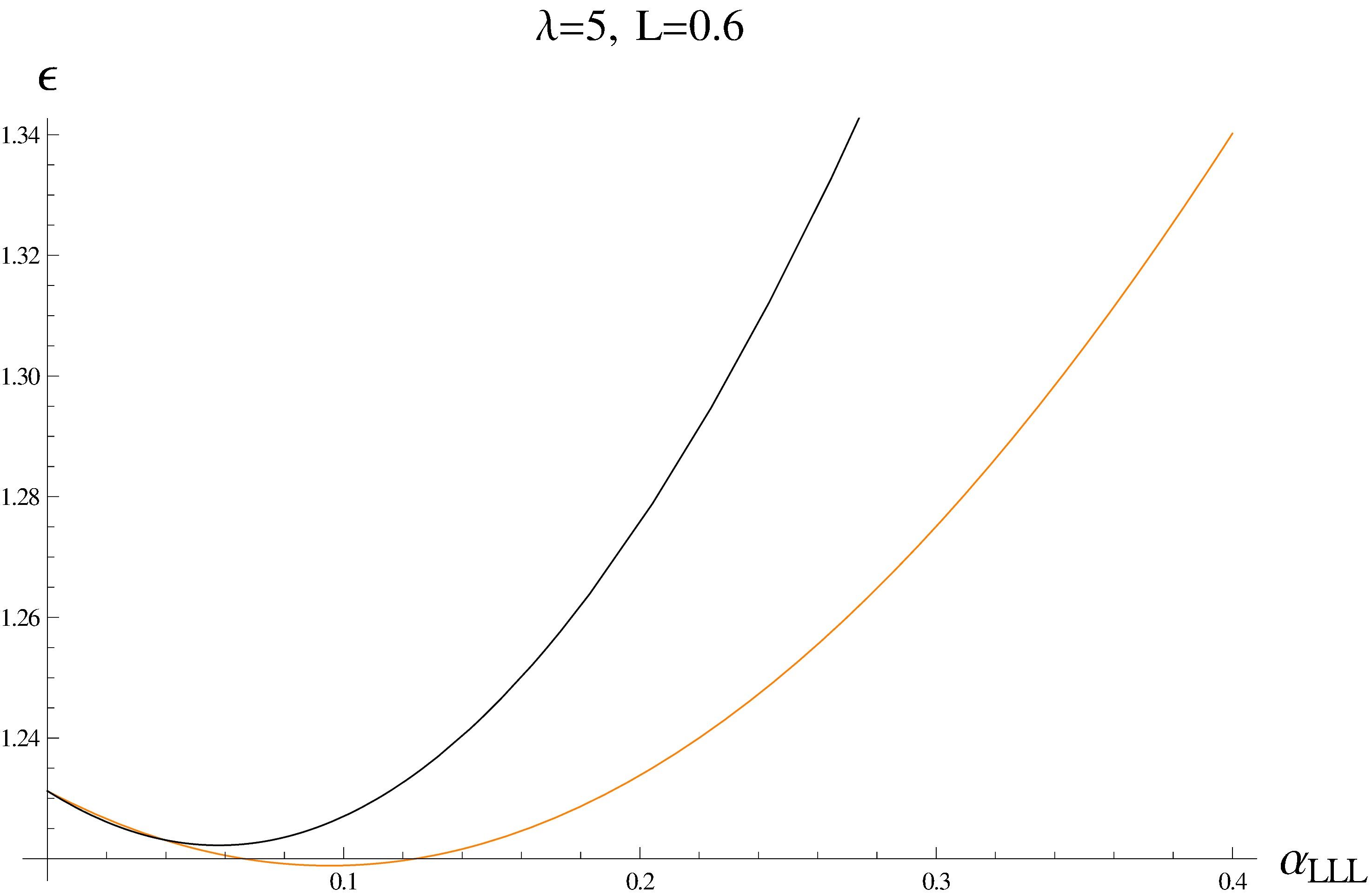}
 \caption{LLL is the optimal basis for a given non-zero AM $L$. The left panel shows the total energy per particle $\e$ for $\la=5$ and $L/N=0.6$ as a function of the each of the six (relative) occupations $\al_i$ of the non-LLL, while all the rest are kept to zero. Any variation of these occupations increases the total energy of the system. In the right panel we plot, for comparison, the dependence of the energy, for the same total AM, on the occupations of the LLL with $m=2$ (orange/lower line) and $m=3$ (black/upper line). A clear minimum can be seen at a non-zero value of $\al$. All calculations are done at the optimal values of $\s$ for $N=6000$ particles.}
\label{LLLeqBMF}
\end{figure}

To give some more weight and generality to this claim, we have examined the states $|\vec n_{10}\rangle$ which are built over the $M$=10 $\s$-scaled orbitals $\{s, p_+, p_-, d_{2+}, d_0, d_{2-}, f_{3+}, f_{+}, f_{-}, f_{3-}\}$.~
We calculated the energy and minimized it simultaneously with respect to the occupations $\al_i=n_i/N,~i=3,\dots,10$ and $\s$ for given $L>0$ and large $N$. 
We found again -- both analytically in the large-$N$ limit and numerically -- that for all allowed $\la$, any non-zero occupations of the non-LLL orbitals $\{p_-,d_0,d_{2-},f_{+},f_{-},f_{3-}\}$ will only increase the total energy $\e[\n{10}]$. Hence, the occupation of any non-LLL is not energetically favorable and indeed the best mean field, for given L, comprises of LLL only. This is demonstrated in Fig.~\ref{LLLeqBMF}. In the left panel, we plot the total energy per particle of the system as a function of each of the six relative occupations of the orbitals that do \emph{not} belong to the LLL, while the rest five of them are set to zero. In the shown case ($\la=5$, $L=0.6$) any variation of the non-LLL occupation increases the energy. Contrarily, on the right panel, we plot the the energy $\e$ versus the occupations $\al_{\text{LLL}}$, with quantum numbers $l=m=2$ and $l=m=3$ respectively. It can be seen clearly that there is a minimum of the energy at a non-zero value of any of the two $\al_{\text{LLL}}$.

\paragraph{Ground state for given L.}
Having found that indeed the BMF is built over the LLL orbitals solely, we consider hereafter permanents of Eq.~(\ref{permanent}) built over LLL only [Eq.~(\ref{LLL})]. With this choice, i.e., $m=l$ and hence using one index $m$ only for each orbital $\phi_m$, the energy functional of Eq.~(\ref{Energies}) becomes:
\begin{align}
\e_{\text{LLL}} &  =  (1+\Lper) h_{0} + \frac{\lao}{2}w_{0,0}\left(\Lper^2/2-1-\frac{\Lper-2}{2N}\right)  +   \nonumber \\ 
&+ \frac \lao 2  \sum_{m}\left[ \left(2-\Lper~m +\frac{m-2}{2N}  \right)w_{0,0}+ 4 (\Lper-1)w_{0,m}- 4 \Lper w_{1,m}\right] \al_{m} +   \nonumber  \\
 & +  \frac \lao 2 \sum_{m,m'} \left(\mathcal K_{m,m'}^+ - \mathcal K_{m,m'}^- \right) \al_{m} \al_{m'}, 
 \label{EnergiesLLL}
\end{align}
and the matrix elements now take on the explicit form:
\begin{eqnarray}
h_{i} = (1+i) \frac{1+\s^4}{2 \s^2} \text{~~and~~} w_{i,j}=\frac{(i+j)!}{2^{1+i+j}  i!j!} \frac{1}{\pi \s^2}.
\label{Energy_LLL}
\end{eqnarray}
Our task now is to find this set of parameters $\{n_i,\s\}$ that for a given $L$, minimizes the \emph{total} energy per particle $\e$.
We have examined and compared the energies of all different possible Fock states built over $M=13$ LLL orbitals, with OAM $m=0,\dots,12$, for a particle number up to $N=18$. Interestingly, we found that above some critical value $L_c$ for the AM the optimal occupations, i.e., the distribution of occupations that minimizes the energy, is given by:
\begin{equation}
 n_0=N-2,~n_1=1,~n_m=\delta_{m,L-1},~m=2,\dots, M = L,
\label{BMFdistri}
\end{equation}
where $\delta_{i,j}$ is the usual Kronecker delta. The same state in a Fock representation reads:
\begin{equation}
|N-2,1,0, \dots,0,1,0,\dots \rangle,
\label{2D_optimal_distri}
\end{equation}
i.e., only the $m=0$, $m=1$ and $m=L-1$ orbitals are populated. We found that this is the optimal distribution of occupations, independent of $L$, as long as this is larger than the approximate value\footnote{Precisely, this critical value is the solution of $L_c^2-L_c-4 N+4+2^{3-L_c}(L_c+2N-4)=0$.} $L_c \simeq 2\sqrt{N}$. For values lower than $L_c$ either the permanent $|N-L,L,0,\dots\rangle$ or the permanent $|N-(L-1),L-2,1,0,\dots\rangle$ are the optimal distributions, depending on the value of $L<L_c$.

There is a simple reasoning why such an unexpected distribution of the bosons among three orbitals only is found to be optimal. Both the prefactors of $\al_m$ as well as that of $\al_m \al_m'$ in Eq.~(\ref{EnergiesLLL}) admit a maximum at $m=M-1$. In other words, the interaction energy is minimized when the `furthest' orbital is occupied. Due to the attraction, the bosons like to sit close to each other, even in the presence of AM. By exciting only one or two bosons in orbitals with the appropriate OAM, the system achieves the desired non-zero AM L at the lowest energetical cost possible.
So, for a given AM $L$, one boson occupying the orbital with OAM $m=L$ is expected to make up the energetically preferable configuration. 
Assuming\footnote{So that the system has the minimum required amount of quanta of AM.} here and hereafter that $L>2$, the energy of such a configuration, as can be directly derived from Eq.~\eqref{En_Phi}, is $\e_{\text{e}}=h_{0}(\Lper +1) + \frac{\lo}{N}\left[\frac{(N-1)(N-2)}{2} w_{0,0}+2 (N-1) w_{L,0}\right]$. However, one can show that if the system excites two bosons, instead of one, to the $m=1$ and $m=L-1$ orbitals the resulting energy will be lower than the previous case. This additional lowering of the energy comes from the \emph{exchange energy} [included in the last term of Eq.~(\ref{EnergiesLLL})] between the two fragments, $\phi_{m=1}$ and $\phi_{m=L-1}$. The energy is now given by
$\e_{\text{BMF}}=h_{0}(\Lper +1) + \frac{\lo}{N} \left[ \frac{(N-2)(N-3)}{2} w_{0,0}+2 (N-2) w_{0,1}+2 (N-2) w_{0,L-1}+2 w_{L-1,1} \right] $ and is indeed the ground state energy for some given $L$.
Substituting the matrix elements in the last expression of the energy we get finally:
\begin{equation}
\e_{int}=- \la \frac{w_{0,0}}{2 N} \left(N-2+2^{2-L}\frac{2N+L-4}{N-1}\right), 
\end{equation}
with $h_{0}=\frac{1+\s^4}{2 \s^2}$ and $w_{0,0}=1/(2 \pi \s^2)$.
We minimize the total energy
\begin{equation}
\e=E/N = \e_0+\e_{int},
\label{etotal}
\end{equation}
where
\begin{equation}
\e_0=\left(1+\Lper \right) h_{0}(\s)
\end{equation}
with respect to $\s$ to arrive at the expression for the optimal energy of an attractive system with a given number of quanta of AM $L=N \Lper$. To keep the clarity in presentation we give here only the expression in the limiting case where $N \gg 1$ and $\Lper=L/N$ is fixed, while the full expression can be found in Appendix~\ref{appen2D}. This reads:
\begin{equation}
 \e_{int}= -\frac{\la}{4\pi \sqrt{1-\frac{\la}{2\pi(\mathcal L+1)}}}.
\label{EnL}
\end{equation}
And the optimal value for the parameter $\sigma$, i.e. the optimal width of the orbitals as a function of the interaction strength and the AM is given by:
\begin{equation}
 \s_0=\left(1-\frac{\la}{2\pi(\Lper+1)}\right)^{-1/4},
\label{s0LLL}
\end{equation}
also in the large-$N$ limit.
We arrive here at a simple expression for the energy and the single-particle states of the moderately and strongly\footnote{Relatively strong $\lambda_0$, of course, so long as the condensate is non-collapsed.} attractive system, with $L=\Lper N$ quanta of angular momentum. From Eq.~(\ref{EnL}) one immediately derives the asympotic relation for $\la\ll 1$ or, equivalently, for large $\Lper$. This reads:
\begin{equation}
\e_{int}=-\frac{\la}{4\pi}
 \label{EnLasym}
\end{equation}
and coincides with the expression given in Refs. \cite{Wilkin1998, Jackson2008}. What we see is that the energy given in the above references is the large-$N$, low-$\la$ limit of Eq.~(\ref{EnL}). Moreover, for large $N$, the energy of Eq.~(\ref{EnL}) is \emph{always} lower than the asymptotic expression $-\la/4\pi$, since it takes into account corrections of finite interaction strength $\la$ beyond first order.

Finally, the total energy per particle, in the large-$N$ limit, reads:
\begin{equation}
 \e=(\Lper+1) \sqrt{1 - \frac{\la}{2\pi (\Lper+1)}}.
\label{etotal2}
\end{equation}
The same expression for small $N$ is shown in the Appendix \ref{appen2D}. It is interesting to note that the resulting optimized energy, as given above, does \emph{not} equal the sum of the non-interacting plus the interaction energy.
They are rather connected through the relation 
\begin{equation}\frac{\partial \e_{total}}{\partial \la}=\frac 1 \la \e_{int}.
\end{equation}
 This nonlinearity stems directly from the optimized orbitals; the interaction will change the shape of the orbitals and this will in turn alter the kinetic and potential energies.

\begin{figure}
\centering
 \includegraphics[scale=0.75]{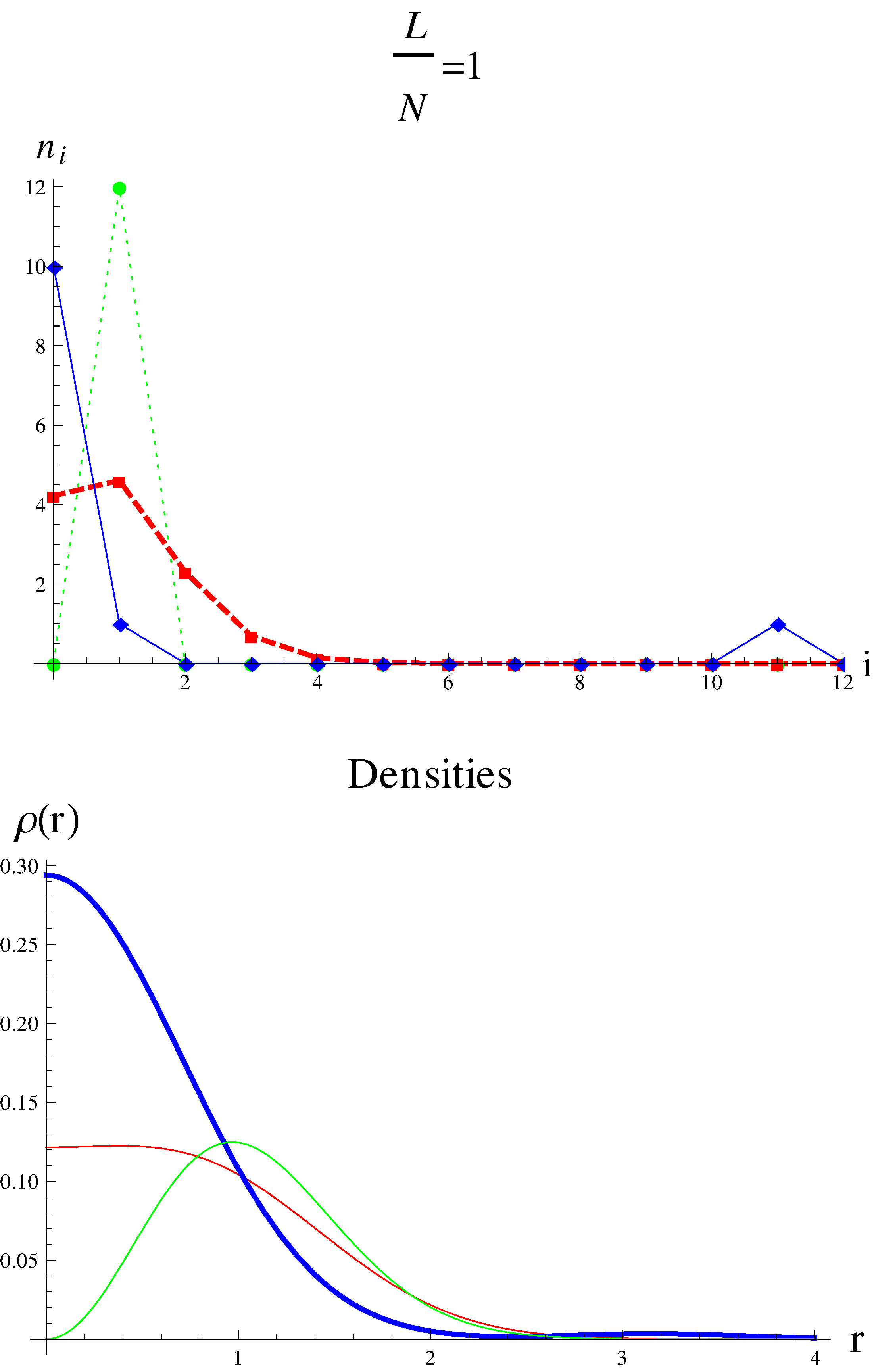}
\caption{Distributions of the occupation numbers (upper panel) and density $\rho_1(r)$, i.e., diagonal of the single-particle reduced density matrix (lower panel) for three different states, all with $\Lper=1$ and $N=12$. The blue line corresponds to the ground state of Eq.~\eqref{2D_optimal_distri}, the red line to the ground state found in Ref.~\cite{Wilkin1998} and the green to the vortex state [see Eq.~\eqref{evor}].  \label{Densities}}
\end{figure}

\paragraph{Quantized vortices.}
A well-studied rotating collective excitation of the quantum gas is the vortex state (see for instance the review of Ref.~\cite*{Fetter2010, Fetter2010b} and references therein). A quantized vortex is the coherent state where all particles of the system are in the excited orbital $\phi_m \sim r^m e^{r^2/(2\s^2) + i m\theta}$ with some vorticity $m\in \mathbb N$. Here again, $\s$ is the scaling parameter. However, a vortex is a highly excited state of the attractive gas with given total angular momentum $L=m\ N$, as can be seen from the comparison of energies of the above-found ground state and the vortex. Its $\s$-optimized energy is easily found to be:
\begin{equation}
 \e_{\text{VOR}}=(\Lper+1) \sqrt{1-\frac{\lambda~ \G(\Lper+\frac 1 2) }{2\pi^{3/2}(\Lper+1) \Lper!}},
 \label{evor}
\end{equation}
where $\G(\dots)$ is the Gamma function.
This energy, compared to that of Eq.~\eqref{etotal2} is always higher.
The vortex state implies a `hole' in the density of the gas and hence - considering the attractive nature of the interaction - is energetically expensive. The distributions and densities of the $m=1$ vortex and of the ground state of Eq.~\eqref{2D_optimal_distri} are compared in Fig.~\ref{Densities}.

\section{Stability of the ground states \label{2D_critical_la}}
Next, we calculate the stability of the above found ground states for some given AM $L$ (or $\Lper=L/N$) found above. In other words, we are interested in the maximum or critical value of the interaction parameter $\la=|\lo|(N-1)$ such that the condensate exists in a non-collapsed state. This is estimated as the maximum value of $\la_c$ such that there is a well defined \emph{global} minimum of the energy as a function of the scaling parameter $\s$ that determines the width of the Gaussian profiles of the orbitals (scaled LLLs). We calculate this value by setting to zero the first and second derivatives of the energy $\e(\s)$ of Eq.~\eqref{etotal} with respect to $\s$. We arrive at the expression:
\begin{equation}
 \la_c = \frac{2^{L+1} (N-1) (N+L) \pi}{4L+[8+2^L(N-1)] (N-2)},
\label{laL}
\end{equation}
which for $N\gg1$ yields:
\begin{equation}
 \la_c \simeq 2 (\Lper+1)\pi = (\Lper+1) \la^{GP},
\label{laLasym}
\end{equation}
where $\la^{GP}=\la_c(\Lper=0)|_{N\gg 1} = 2 \pi$ is the critical $\la$ of the GP condensed ground state with zero AM.
So, as long as $N$ is sufficiently large, practically above a few hundreds of particles, the critical interaction parameter $\la_c$ increases linearly with the AM $\Lper$.
Equations~(\ref{laL}) and (\ref{laLasym}), together with Eqs.~(\ref{EnL}) and \eqref{etotal2} are the main results of this work.
It should be noted that the corresponding critical value for $\la$ of a vortex state of vorticity $\Lper$ is higher than the one given above for the GS. Precisely, from Eq.~\eqref{evor} we immediately obtain $\la_c^{\text{VOR}}=2 \pi \frac{\sqrt{\pi} (\Lper+1)\Lper!}{\Gamma(\Lper + \frac 1 2)}$. The fact that $\la_c^{\text{VOR}} > \la_c$ comes to no surprise, since a vortex state is a highly excited state of the attractive system.

\section{Comparisons with known results \label{MBcomparison}}
\begin{figure}[ht!]
 \centering
\includegraphics[scale=0.75]{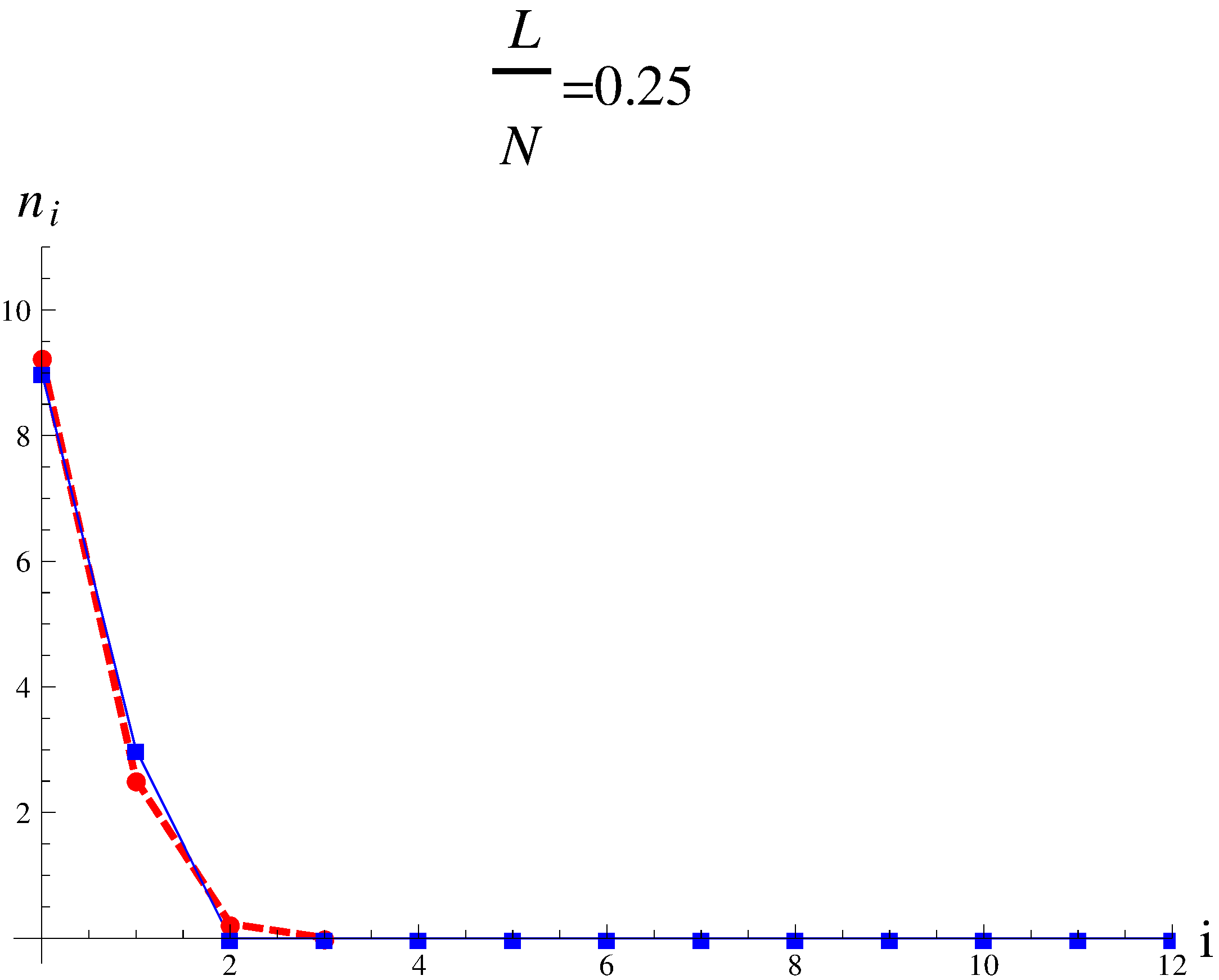}
\includegraphics[scale=0.75]{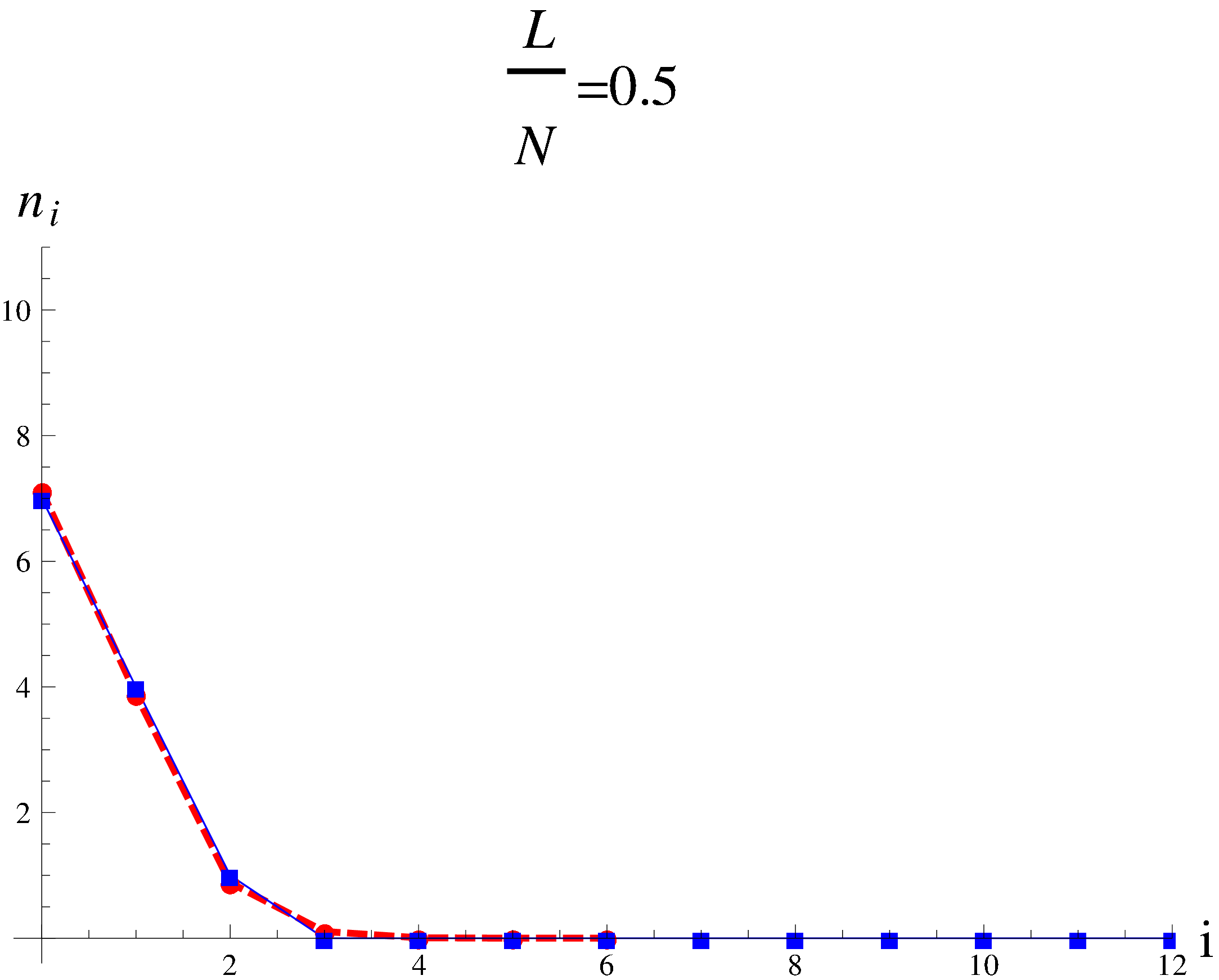}
\includegraphics[scale=0.75]{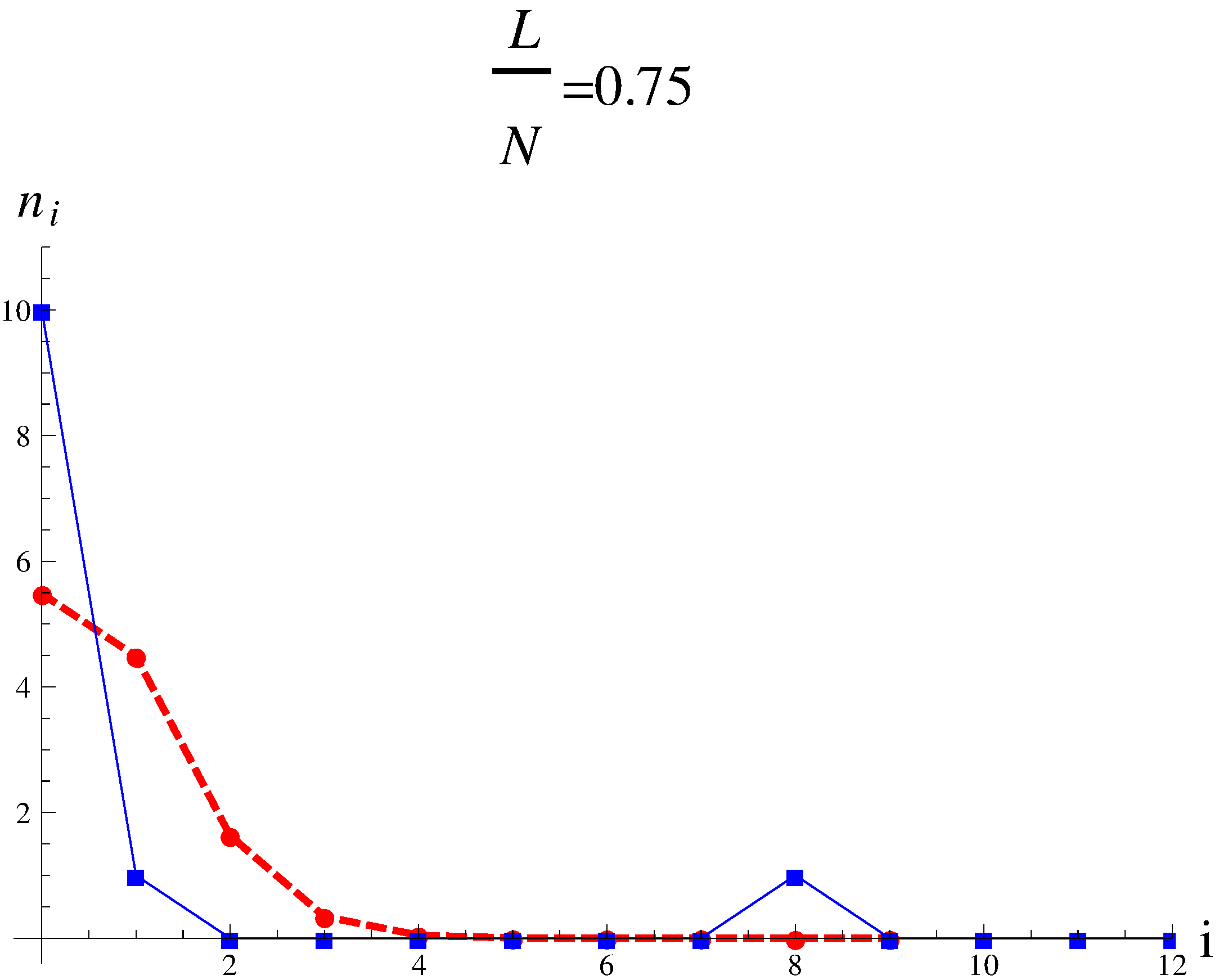}
\includegraphics[scale=0.75]{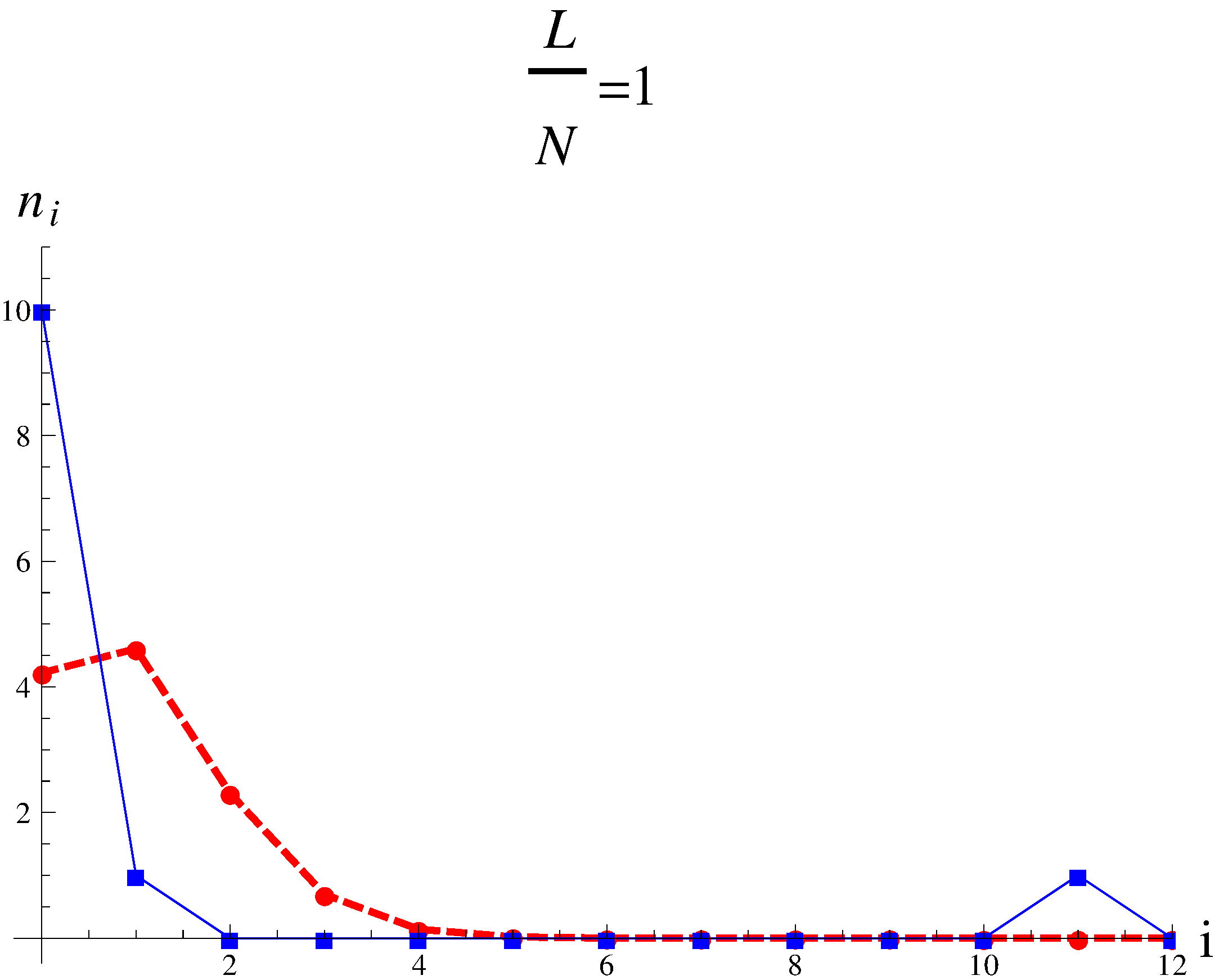}
\caption{Occupation numbers of the ground state for different values of $\Lper$, as found within the BMF theory (blue) and as given in Wilkin et al. \cite{Wilkin1998} (red-dashed). The agreement is good for an approximate value of $\Lper \lesssim 0.5$. Above this value the two distributions take on a completely different appearance, even if the energies of the configurations are almost equal. For instance, for a weak interaction $\la=0.01$ the energy difference of the two is $\Delta E\sim 10^{-4}$. The number of particles here is $N=12$.
\label{occupations}}
\end{figure}

Finally, we compare the energies and occupations obtained in the preceding sections with known results obtained at the MB level. In the work of Wilkin et al. \cite{Wilkin1998}, as well as that of Jackson et al. \cite{Jackson2008} the following result is given for the total energy of a weakly attractive system:
\begin{equation}
 \e_W=\Lper +1 - \frac{\la}{4\pi}.
 \label{EJ}
\end{equation}
It is interesting that this result is obtained both within a MB approach \cite{Wilkin1998} and a MF ansatz \cite{Jackson2008}. Wilkin et al. \cite{Wilkin1998} start by writing the (not normalized) solution of the problem as
\begin{equation}
\psi_W =  r_c^L  e^{-\sum_i^N r_i^2/2 },
\label{PsiWilkin}
\end{equation}
where $r_c$ is the centre of mass coordinate, and find that the natural orbitals $\phi_m$ of the system are the LLL states. That is, single-particle states $\phi_m \propto r e^{-r^2/2 + i m\theta}$, which unlike our calculations are not scaled. The respective natural occupations, for $N$ particles and $L$ total AM, are found to be \cite{Wilkin1998}:
\begin{equation}
 \rho_m=\frac{(N-1)^{L-m} L!}{N^L (L-m)! m!}.
 \label{rhoW}
\end{equation}
The interaction energy of such a configuration equals the interaction energy of the non-rotating system, i.e., $\e_{\text{int,W}}=-\frac{\la}{4\pi}$. In the more recent treatment of Ref.~\cite{Jackson2008} the authors built up a GP ansatz out of the fragments $\phi_m$ and their occupations found in Ref.~\cite{Wilkin1998}. Specifically, they expressed the GS of the gas with total AM $\Lper$ as $\psi_J=\sum c_i \phi_i$, where $c_i$ are the large-$N$ and large-$L$ limits of the occupations $\rho_m$ of Eq.~\eqref{rhoW} and the orbital-basis $\{\phi_i\}$ is again the LLL. The energy thus obtained exactly equals that of Eq.~\eqref{EJ}.
We immediately see that the energy found in both of the above approaches is the same as the vanishing-$\la$ limit of Eq.~(\ref{EnL}). Hence, we are able to reproduce the known result and, moreover, give higher-order corrections due to finite interaction strength $\la$.
In Fig.~\ref{occupations} we plot the occupations of the LLL states for $N=12$ and different values of $\Lper$, as calculated in our BMF approach and compare them to those of Eq.~\eqref{rhoW}.

\newpage
\section{Discussion and conclusions \label{conclusions}} 
We have theoretically demonstrated herein and in agreement with earlier work, that the attractive gas with given angular momentum per particle prefers to fragment over a finite set of single-particle states (orbitals). To determine these states we used the scaled Gaussian functions with certain angular momentum that form the so-called lowest Landau levels (LLL) and, moreover, a variational argument was given to justify the use of LLL. Based on these states a multi-orbital mean field was constructed to calculate the total energy, the fragmentation and the stability of the ground states. 

We arrived at the result that, for given total AM $L$ larger than a critical value, only two of the excited shape-optimized orbitals are populated and these carry all the angular momentum of the system. Specifically, for $L\gtrsim 2\sqrt N$ it is only one particle with a single quantum of angular momentum and another one with the rest $L-1$ [see Eq.~\eqref{2D_optimal_distri}]. Our results are valid for weak but also moderately large interaction strength $\la$. The inclusion of the $\s$-parameter gives an extra flexibility to our ansatz [Eq.~(\ref{LLL})] and allows for a description of the collapse. 
The accuracy of this ansatz has been checked numerically. The ground state of the system has the appearance of a Gaussian distribution for all allowed interaction strengths and the agreement with the employed variational ansatz has been found to be satisfactory. While the predicted $\lambda_c$ is shifted to higher values in the variational treatment, qualitatively (and quantitatively up to a finite moderately large value of $\lambda$) the numerical and analytical approaches agree (see also Ref.~\cite{Tsatsos2010}).
With that, we calculated the energy of the ground state possessing some finite angular momentum per particle $\Lper$ as a function of $\Lper$ and $\la$. This finding constitutes a generalization of the previously known result: indeed, from a first order expansion for small $\la$ we got back the relation for the energy as first presented by Wilkin et al. and later on by Jackson et al., at the many-body \cite{Wilkin1998} and Gross-Pitaevskii \cite{Jackson2008} levels respectively. 

The inclusion of scaled orbitals in our variational multi-orbital approach was a crucial step and allowed us to calculate the finite-$\lambda$ corrections to the stability and energy of the 2D attractive gas. Due to this correction the total energy can drop lower than that that predicted in the many-body analysis of Ref.~\cite{Wilkin1998}. The deviation of our predictions for the distribution of the occupations, above a critical AM, from that predicted in Reference \cite{Wilkin1998} should be attributed to the different approaches followed. Herein, we followed a MF approach with a truncated (to M=13 orbitals) Hilbert space. 
A complete  study of the 2D attractive gas requires though a (numerically) exact analysis for all allowed $\lambda$. 
Such a study would shed more light on the structure of the many-body (ground or excited) states, the exact shape of the single-particle states and their respective occupations as well as on the quantum fluctuations, that are, by definition, left out in a mean field description and are known to grow large for growing $\lambda$ \cite{Tsatsos2010}. It would thus be decided whether the occupations of the LLL for given AM, found here, persist at the MB level. 
Such a treatment can be accomplished, e.g., by the usage of the Multi-Configurational Time-Dependent Hartree method for Bosons (abbreviated as MCTDHB) \cite{MCHB,RoleOfExcited}, which is the subject of forthcoming work.

Concerning many-body computations, the choice of an adequate orbital-basis is crucial. In other words, what is the optimal number of orbitals that should be included in a many-body study so that the calculation is converged? Although, a thorough discussion of this fundamental numerical issue goes beyond the scope of the present work we  comment on the use of Dirac delta distribution as the two-body interaction pseudopotential in many-body (i.e., beyond mean field) theoretical approaches.
It was recently shown \cite{Rosti} that the use of a (non-regularized) Dirac delta pseudopotential in a system of two interacting ultracold bosons in two dimensions leads to very slow convergence of the ground state energy to that of the non-interacting ground state energy, for any value of the repulsive interaction strength, and to minus infinity for any value of the attractive interaction strength. Thus, the truncation of Hilbert space becomes problematic. In such a case, a narrow Gaussian model for the two-body interaction is a suitable choice for the many-body study \cite{Rosti}. 

Lastly, we note that the present findings, together with these of Ref.~\cite{Tsatsos2011} could play a significant role in vortex engineering in ultracold atomic gases. The modulation or change in sign of the scattering length could prove helpful in controlling the way the cloud absorbs angular momentum from its environment so that vortex clusters and giants vortices can form and be manipulated. Such engineering could find application in exploring the more complex turbulent atomic gas with large number of vortices, relatively recently achieved in the laboratory \cite{Henn2009}.

\begin{acknowledgments}
A great part of the present work has been carried out in the Theoretical Chemistry group of Heidelberg University. The author is indebted to L. S. Cederbaum, O. E. Alon and A. I. Streltsov for providing guidance and inspiring the present work. V. Romero-Roch\'in and the Institute of Physics of the National Autonomous University of Mexico are greatly acknowledged for providing kind hospitality. Financial support from the HGSFP, the STIBET fund of Heidelberg University as well as the DFG grant CE 10/52-1 is acknowledged.
\end{acknowledgments}

\appendix
\section{Full expression for the energy of the 2D gas \label{appen2D}}
We give the full expressions of the total and interaction energy for a 2D attractive system, as calculated within the BMF theory.
Total energy, optimized for $\s$ for any $L>2$, $N$ and $\la$:
\begin{equation}
 \e_{f}=(\Lper+1) \sqrt{1 -  \frac{\la}{2\pi (\Lper+1)} A(L,N)},
\label{2Dfullenergy}
\end{equation}
where 
\begin{equation}
 A(L,N)=1-  \frac 2 N + 2^{2-L} \frac{2 N + L - 4}{N (N-1)}
\end{equation}
and $\Lper = L/N$.
For $N\rightarrow \infty$ we get $A\rightarrow 1$ and $\e_f \rightarrow \e$, i.e., the above expression reduces to the energy of Eq.~\eqref{etotal2}.
The optimum $\sigma$, i.e., the value $\s_0$ where the total energy obtains a minimum is:
\begin{equation}
\s_0=\sqrt[4]{1-\frac{\la}{2\pi (\Lper+1)} A(L,N)}.
\label{optimumsigma2D}
\end{equation}
The optimized interaction energy reads:
\begin{equation}
\e_{\text{int},f} = - \frac{\la A(L,N)}{4 \pi \sqrt{1-\frac{\la A(L,N)}{2 \pi (\Lper +1)}}}	
\label{2Dfullenergyint}
\end{equation}
Similarly, this is the general expression for the energy as given the Eq.~\eqref{EnL} but without having taken the large-$N$ limit. Note that the above represents the ground state energy as long as $L>L_c$, as explained in the main text.

\end{document}